\documentclass[aps,pra,groupedaddress,twocolumn,longbibliography,superscriptaddress]{revtex4}

\usepackage{cancel}
\usepackage{graphicx}
\usepackage{amssymb, amsmath}
\usepackage{bbold}
\usepackage{float}
\usepackage{multirow}
\usepackage{nicefrac}
\usepackage{xcolor}
\usepackage{textcomp}

\usepackage{pifont}

\newcommand{\mean}[1]{\left\langle #1 \right\rangle}

\newcommand{\ket}[1]{\vert #1 \rangle}
\newcommand{\ketbra}[2]{\left\vert #1 \middle\rangle\middle\langle #2 \right\vert}
\newcommand{\n}{\left\langle n \right\rangle}
\newcommand{\nn}[1]{\langle n^{(#1)} \rangle}
\newcommand{\al}{|\alpha|}
\newcommand{\Poisson}{\textnormal{Poisson}}

\usepackage[normalem]{ulem}

\begin{document}

\title{Differences and similarities between lasing and multiple-photon subtracted states}

\author{T.~Lettau}
\affiliation{Institute of Condensed Matter Theory and Solid State Optics, Abbe Center of Photonics, Friedrich-Schiller-Universit{\"a}t Jena, Max-Wien-Platz 1, 07743 Jena, Germany}

\author{H.A.M.~Leymann}
\email{ham.leymann@gmail.com}
\affiliation{INO-CNR BEC Center and Dipartimento di Fisica, Universita di Trento, I-38123 Povo, Italy}

\date{\today}

\begin{abstract}
We examine the effect that the subtraction of multiple photons has on the statistical characteristics of a light field.
In particular, we are interested in the question whether an initial state transforms into a lasing state, i.e.,~a (phase diffused) coherent state, after infinitely many photon subtractions.
This question is discussed in terms of the Glauber P-representation $P(\alpha)$, the photon number distribution $P[n]$, and the experimentally relevant autocorrelation functions $g^{(m)}$.
We show that a thermal state does not converge to a lasing state, although all of its autocorrelation functions at zero delay time converge to one.
This contradiction is resolved by the analysis of the involved limits, and a general criterion for an initial state to reach at least such a pseudo-lasing state ($g^{(m)}\to 1$) is derived, revealing that they can be generated from a large class of initial states.
\end{abstract}

\maketitle
\section{Introduction}
\label{sec:intro}
The effect of photon subtraction (and addition) has been studied in recent years in the context of probing commutation rules \cite{parigi_probing_2007}, quantum information processing \cite{ra_tomography_2017} and quantum key distribution \cite{barnett_quantum_2009,lim_longer_2018,ma_continuous-variable_2018,zhao_continuous-variable_2018},
meteorology with photon subtracted Gaussian states, precise phase measurements \cite{braun_precision_2014,rafsanjani_quantum-enhanced_2017}, quantum state engineering \cite{sperling_quantum_2014}, and general thermodynamic considerations like Maxwell's demon \cite{vidrighin_photonic_2016}.
A textbook discussion of photon subtracted states (PSSs) can be found in \cite{agarwal_quantum_2013} and older theoretical studies of PSSs \cite{agarwal_negative_1992} are complemented by a recent overview article on the statistics of photon subtracted and added states \cite{barnett_statistics_2018}.
There are many experimental realizations of single- \cite{kim_nonclassicality_2005, ourjoumtsev_generating_2006, biswas_nonclassicality_2007} and multi-PSSs \cite{fiurasek_engineering_2009, bogdanov_multiphoton_2017}.
In most of these experiments the setup for the photon subtraction is based on a weakly
reflecting beam splitter and photon detectors.
This realization of photon subtraction is a probabilistic process and the chances of success decrease significantly with the number of subtractions.
For details on the success probability and the influence of the detector efficiency, we refer the reader to \cite{sperling_quantum_2014,allevi_reliable_2010,barnett_statistics_2018,bogdanov_multiphoton_2017}.
Experiments on thermal states with up to ten subtractions, along with a theory that is derived from the generating function for the photon number distribution, can be found in \cite{bogdanov_multiphoton_2017,bogdanov_study_2016}.

\begin{figure}[!hbt]
  \centering
  \includegraphics[width=0.99\columnwidth]{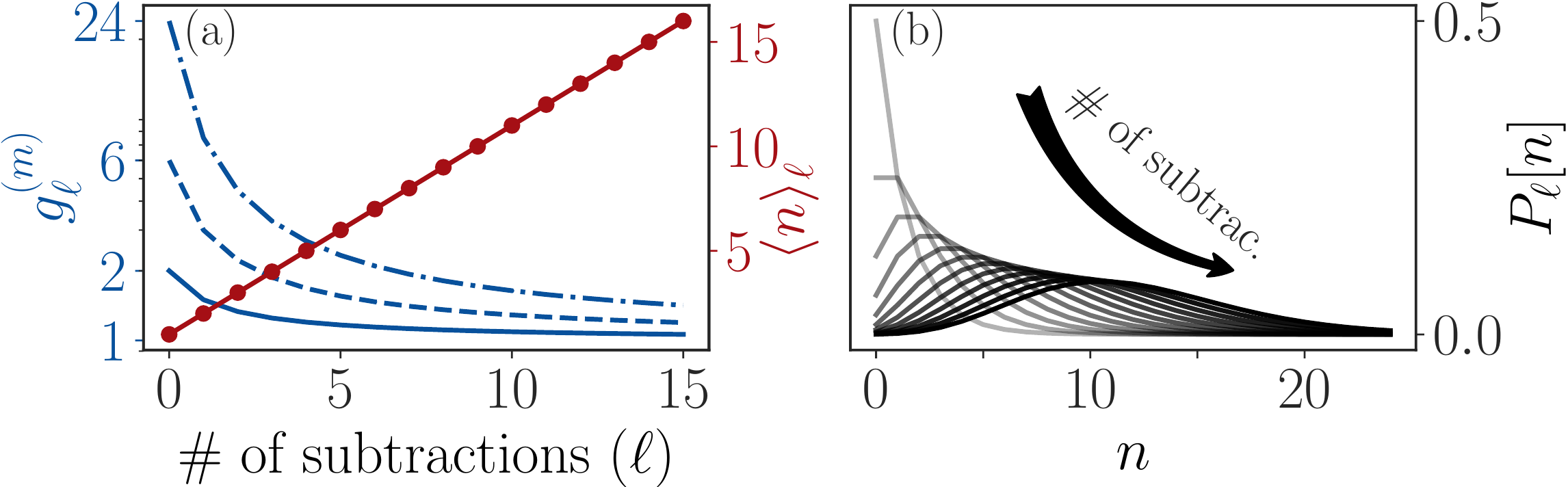}
  \caption{(a): Numerical results for the autocorrelation functions $g^{(m)}$ (measured on the left y-axis) for $m=2$ (blue, solid), $m=3$ (blue, dashed) and $m=4$ (blue, dot-dashed) and mean $\n$ (red, big dots, (measured on the right y-axis)) of an initially thermal $\ell$-PSS.
(b): Photon number distributions of the same states for $\ell \in[0\dots 12]$ photon subtractions.}
  \label{fig:numericgmpn}
\end{figure}

Photon subtraction is one of the most fundamental processes in quantum optics and its effect on the initial density operator $\rho_0$ can be described by the photon annihilation operator $a$.
The density operator of the $\ell$-PSS for a single optical mode, i.e.,~a state after the subtraction of $\ell$ photons is given by \cite{barnett_statistics_2018}
\begin{align}
  \rho_{\ell}=\frac{a^\ell \rho_0 a^{\dagger \ell}}{\textnormal{Tr}[a^\ell \rho_0 a^{\dagger \ell}]}.
  \label{eq:densetyofPSS}
\end{align}
Using this expression, one can study the effects of photon subtraction on the photon statistics directly by using a numerical quantum optics tools like QuTiP \cite{johansson_qutip_2013}.
Figure~\ref{fig:numericgmpn} shows numerical results for the autocorrelation functions $g^{(m)}=\nicefrac{\mean{a^{\dagger m}a^{ m}}}{\mean{a^{\dagger }a}}$ \cite{glauber_quantum_1963},
the mean photon number $\n$ (Fig.~\ref{fig:numericgmpn} (a)),
and the photon number distributions $P[n]=\langle n\vert\rho\vert n\rangle$ (Fig.~\ref{fig:numericgmpn} (b)) of the $\ell$-PSSs starting from a thermal state with $\n_0=1$.
As already shown in \cite{bogdanov_multiphoton_2017}, not only the intensity of the $\ell$-PSSs increases linearly with $\ell$ but also the second- and higher-order  autocorrelation functions at zero delay time $g^{(m)}$ decay from $m!$ to one for increasing $\ell$ (see Fig.~\ref{fig:numericgmpn} (a)).
The second- and higher-order autocorrelation functions play an important role in the study of optical coherence \cite{glauber_quantum_1963}
and are frequently used to characterize the statistical properties of the light field, e.g.,~in the context of
single-photon sources \cite{kurtsiefer_stable_2000,eisaman_invited_2011} and to identify lasing operation in micro lasers \cite{ulrich_photon_2007}.
Since a lasing state is regarded as a phase diffused coherent state with a Poissonian photon number distribution \cite{walls_quantum_2008}, $g^{(m)}$ dropping from $m!$ (thermal) to one (Poisson) is a strong indication for lasing \cite{chow_emission_2014},
when criteria like the intensity jump in the input-output curve \cite{rice_photon_1994} are not present.
Note that we are only considering the coherence properties at zero delay time here (see App.~\ref{app:delaytime} for $g^{(2)}(\tau)$) and not the linewidth or the coherence time \cite{samuel_how_2009}, thus when we refer to a lasing state this is to be understood as a high intensity phase diffused coherent state
\footnote{We are only concerned with states without phase information, thus we omit the specification 'phase diffused' in the text from now on.}.
Analyzing the photon statistics of a light source using the autocorrelation functions is also advantageous since they can be measured by various accessible techniques, e.g.,~directly in a Hanbury Brown and Twiss setup \cite{brown_correlation_1956} or indirectly by photon number resolving techniques \cite{asmann_measuring_2010,schlottmann_exploring_2018}.

Additionally, Fig.~\ref{fig:numericgmpn} (b) shows that the photon number distribution evolves from a broad thermal distribution to a distribution centered around its mean $\n$.
These observations could lead to the tempting and also quite paradoxical conclusion that one can create a high intensity laser state by subtracting a large number of photons from an initially low intensity thermal state
\footnote{Readers who are interested in the paradox that photon subtraction can lead to a higher intensity, are referred to \cite{barnett_statistics_2018}, where this is discussed in therms of post selection and Bayesian reasoning.}.
It is the aim of this article to discuss this issue in detail and to answer the question:
\emph{Can the subtraction of multiple photons create a lasing state?}
In more general terms:
What happens to the photon statistics, when a substantial number of photons is subtracted?
This question becomes relevant in the light of the recent theoretical interest in PSSs \cite{barnett_statistics_2018},
(multi) photon subtraction experiments eg.~\cite{fedorov_quantum_2015,bogdanov_multiphoton_2017,avosopiants_non-gaussianity_2018} and in the context of the realization of probabilistic amplifiers \cite{kim_quantum_2012,chrzanowski_measurement-based_2014,zavatta_high-fidelity_2011}.
To provide an answer to this question that goes beyond the numerical hints provided in Fig.~\ref{fig:numericgmpn}, we study the P-representation $P(\alpha)$ \cite{glauber_coherent_1963}, the photon number distribution $P[n]$, and the autocorrelation functions $g^{(m)}$ of the $\ell$-PSSs analytically, and compare them to the ones of the  corresponding coherent state.

\section{P-representation of the $\ell$-photon subtracted state}
\label{sec:P-rep}
Since a coherent state is an eigenstate of the annihilation operator, it is instructive to express the density operator of the $\ell$-PSS in the basis of the coherent states by the  function $P(\alpha)$ (P-representation) \cite{cahill_density_1969}.
To obtain $P(\alpha)$ we first rewrite $\rho_\ell$ in a more convenient form
\begin{align}
  \rho_{\ell}=\frac{a^\ell \rho_0 a^{\dagger \ell}}{\nn{\ell}_0}=\frac{a^\ell \rho_0 a^{\dagger \ell}}{g^{(\ell)}_0\n^\ell_0},
  \label{eq:photann}
\end{align}
where we express the $\ell$th-order factorial moment of the initial state
by the $\ell$th-order autocorrelation function ${\nn{\ell}_0=\mean{a^{\dagger m}a^{ m}}_0=g^{(\ell)}_0\n^\ell_0}$.
We use the lower index $\ell$ to express that this quantity is taken from a $\ell$-PSS, e.g.,~$\nn{m}_\ell=\textnormal{Tr}[a^{\dagger m} a^{m}  \rho_{\ell}]$.
In the P-representation
\begin{align}
  \rho=\int P(\alpha)\ketbra{\alpha}{\alpha}d^2\alpha,
\end{align}
annihilation and creation can be expressed as: $a\rho \equiv \alpha P(\alpha)$ and $\rho a^\dagger\equiv \alpha^* P(\alpha)$ \cite{walls_quantum_2008}.
Thus we can express $P_{\ell}(\alpha)$ by multiplying $P_0(\alpha)$ with $\al^{2\ell}/\nn{\ell}_0$
\begin{align}
  P_{\ell}(\alpha)
    =P_0(\alpha)\frac{\al^{2\ell}}{\nn{\ell}_0}
    =P_0(\alpha)\frac{\al^{2\ell}}{g^{(\ell)}_0\n^\ell_0}.
  \label{eq:generalpss}
\end{align}
The P-representation of an initial thermal state is \cite{walls_quantum_2008}
\begin{align}
  P_0(\alpha)=\frac{\exp(-\al^2/\n_0)}{\pi \n_0}
\end{align}
thus the P-representation of $\rho_{\ell}$ is given by
\begin{align}
  P_{\ell}(\alpha)=\frac{\exp(-\al^2/\n_0)}{\pi \n_0}\frac{(\al^2/\n_0)^\ell}{\ell!}=\frac{\Poisson_\ell(\lambda)}{\pi \n},
  \label{eq:poisson_p}
\end{align}
which is a Poisson distribution with $\lambda=\al^2/\n_0$.
Note that here the discrete value $\ell$ is the parameter and $\lambda(\alpha)$ the variable.
Approximating this function by a Gaussian,
\begin{align}
  P_{\ell}(\alpha)
    \approx \frac{\exp\left(-\frac{((\ell+1)\n_0-\al^2)^2}{2(\ell+1)\n_0^2}\right)}{\pi \sqrt{2 \pi (\ell+1)\n_0^2}}
    =\frac{\textnormal{Gauss}_{\mu,\sigma}(\al^2)}{\pi},
  \label{eq:gauss}
\end{align}
centered around $\mu=(\ell+1)\n_0$ and with a width $\sigma^2=(\ell+1)\n_0^2$, which works very well for large mean values (see App.~\ref{app:continuos}), reveals that the center of $P_\ell(\alpha)$ and its width increases linearly with $\ell$.

The P-representation of a coherent state $\ket{\alpha_c}$ is $P_{\textnormal{coh}}(\alpha)=\delta^2(\alpha-\alpha_c)$.
In contrast to the PSS of a thermal state, this state has phase information.
This becomes obvious in polar coordinates
\begin{align}
  P_{\textnormal{coh}}(\alpha)=\delta(\al-|\alpha_c|)\delta(\phi-\phi_c)\al^{-1}.
\end{align}
To compare the P-representation of a coherent state to the one of the PSS we need to diffuse the phase information of the pure coherent state
\begin{align}
  P_{\textnormal{diff coh}}(\alpha)=\delta(\al-|\alpha_c|)/(2\pi\al).
  \label{eq:prepofcoherent}
\end{align}
As we can see from the Gaussian approximation in Eq.~(\ref{eq:gauss}), $P_{\ell}(\alpha)$ does not converge to $P_{\textnormal{diff coh}}(\alpha)$ [Eq.~(\ref{eq:prepofcoherent})].
In general, a state $\rho$ can have different P-representations, since the coherent states form an overcomplete non-orthogonal basis.
Thus showing that two states have different P-representations is not sufficient to show that they represent different states \cite{sperling_characterizing_2016}.
However, in our case $P_{\textnormal{diff coh}}$ and $P_{\ell}(\alpha)$ are both non negative and have only a finite number of $\delta$ singularities.
In this case, we can conclude that $P_{\textnormal{diff coh}}$ and $P_{\ell}(\alpha)$ represent different states (see App.~\ref{app:uniquep}).
Thus, we can already answer our initial question: "Can one create a lasing state by multiple photon subtraction from a thermal state?", by giving the answer "No".
However, there is more to the $\ell$-PSSs.

\subsection{$\ell$-PSSs with scaled intensity}
We can actually create a coherent state, when we scale the initial intensity $\n_0$ in Eq.~(\ref{eq:gauss})
\begin{align}
  \n_0=\n_{\textnormal{F}}/(\ell+1),
  \label{eq:scaling}
\end{align}
such that the resulting mean of the Gaussian $\mu=\n_{\ell}$ is fixed to the value $\n_{\textnormal{F}}$.
In this way we obtain the P-representation
\begin{align}
  \tilde{P_{\ell}}(\alpha)
    \approx \frac{\exp(-\frac{(\n_{\textnormal{F}}-\al^2)^2}{2\varepsilon})}{\pi \sqrt{2 \pi \varepsilon}}\to \delta(\n_{\textnormal{F}}-\al^2)/\pi,
\end{align}
which is one of the standard expressions converging to a delta function
with $\varepsilon=\n_{\textnormal{F}}^2/(\ell+1)\to0$.
Transforming the argument of the $\delta$-function from $\al^2$ to $\al$ we obtain the P-representation for a coherent state $P_{\textnormal{diff coh}}$
\begin{align}
  \lim_{\ell \to \infty}\tilde{P_{\ell}}(\alpha)
    =\delta\left(\n_{\textnormal{F}}^{\nicefrac{1}{2}}-\al\right)/(2\pi\al).
\end{align}
This operation, i.e.,~including the scaling of the initial intensity, is not as meaningful and achievable as the simple photon subtraction, since it requires to prepare a specific initial intensity depending on the number of successful photon subtractions.
Furthermore, decreasing the intensity with $\ell^{-1}$ diminishes the success probability of $\ell$ subtractions considerably \cite{allevi_reliable_2010}.

\section{Photon number distribution}
\label{sec:photnumberdistrb}
From the P-representation of the $\ell$-PSSs we can obtain the photon number distribution
\begin{align}
  P_{\ell}[n]=&\langle n|\rho_{\ell}|n\rangle
    =\int P_{\ell}(\alpha) |\langle n|\alpha\rangle|^2 d^2\alpha\nonumber\\
    =&\int P_{\ell}(\alpha) \frac{\al^{2n}}{n!}\exp(-\al^2) d^2\alpha.
  \label{eq:pln_defi}
\end{align}
Inserting $P_{\ell}(\alpha)$ from Eq.~(\ref{eq:poisson_p}) and using a standard integral (see App.~\ref{app:calculation}), Eq.~(\ref{eq:pln_defi}) results in
\begin{align}
P_{\ell}[n]=\frac{(\ell+n)!}{\ell!\,n!}\frac{1}{(1+\n_0)^{(\ell+1)}} \left(\frac{\n_0}{(1+\n_0)}\right)^n,
\label{eq:negbinom}
\end{align}
which has already been reported, e.g.,~in \cite{barnett_statistics_2018}.
\begin{figure}
  \centering
  \includegraphics[width=\columnwidth]{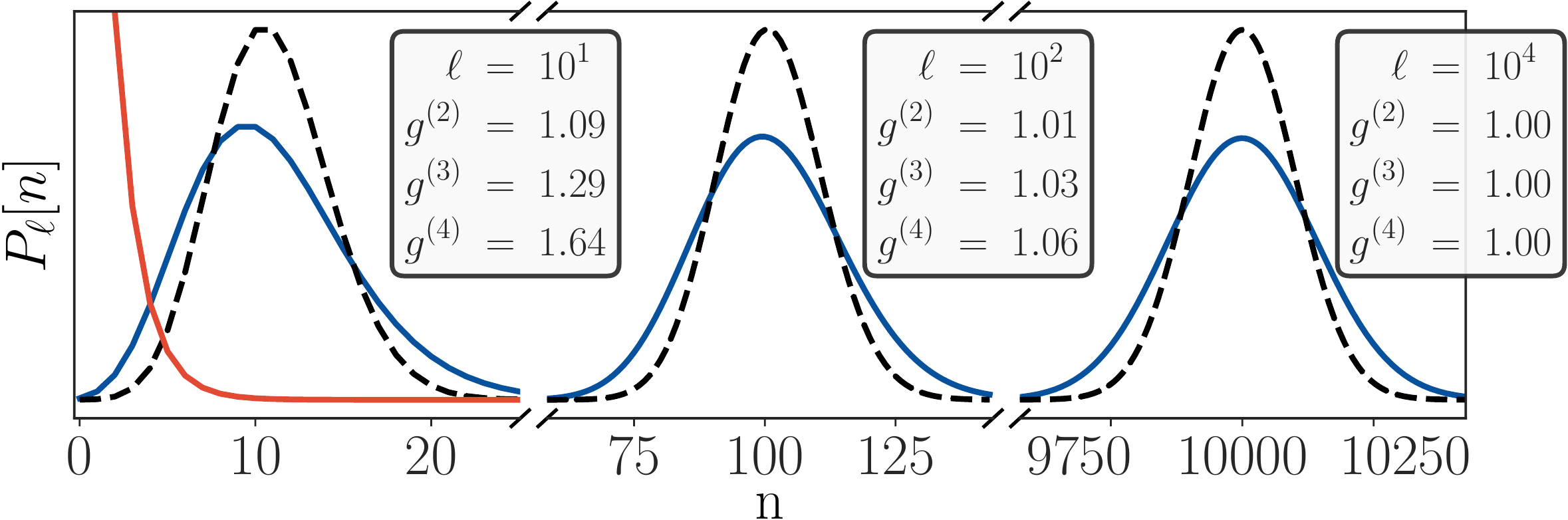}
  \caption{The photon number distribution (blue solid curve) of the $\ell$-PSS after $\ell \in [10, 100, 10000]$ subtractions in comparison to the Poisson distribution with the same mean photon number $\n_0$ (black dashed curve).
  The photon number distribution of the initial thermal state with $\n_0=1$ is depicted on the left by the orange solid curve.
  The boxes show the number of photon subtractions $\ell$ and the corresponding autocorrelations $g^{(m)}$ of order ${m=2,3,4}$.
  Note that the x- and y-axis have different scaling for each $\ell.$}
  \label{fig:poisson_subtract}
\end{figure}
As illustrated in Fig.~\ref{fig:poisson_subtract} this distribution does not converge to a Poisson distribution, which would corresponds to a coherent state, in the limit $\ell\to\infty$.
We see that $P_{\ell}[n]$ is always considerably broader than the corresponding Poisson distribution and that the shape difference of both distributions stabilizes.
By comparing the distribution $P_{\ell}[n]$ [Eq.~(\ref{eq:negbinom})] to the compound Poisson distribution presented in Ref.~\cite{bogdanov_multiphoton_2017}, we can show analytically that this is true.
The compound Poisson distribution is a generalization of the Poisson distribution with two parameters: the mean $\mu$ and the coherence parameter $a$.
For $a=1$ this distribution is a thermal distribution and for $a\to\infty$ it converges to a Poisson distribution.

To turn Eq.~(\ref{eq:negbinom}) into the compound Poisson distribution with an independent coherence parameter that grows with $\ell$, the initial intensity $\n_0$ needs to be scaled by $(\ell+1)^{-1}$ [Eq.~(\ref{eq:scaling})].
The new photon number distribution
\begin{align}
\tilde{P}_{\ell}[n]=\frac{(\ell+n)! }{\ell!(\ell+1)^n}\,\frac{\n_{\textnormal{F}}^n}{ n!} \,\left(\frac{1}{1+\frac{\n_{\textnormal{F}}}{\ell+1}}\right)^{n+\ell+1},
\end{align}
with the scaled intensity now has a fixed mean ${\mu=\n_{\textnormal{F}}}$ photon number and a coherence parameter ${a=\ell+1}$. $\tilde{P}_{\ell}[n]$ is identical to the compound Poisson distribution reported in \cite{bogdanov_multiphoton_2017} and converges to the Poisson distribution for $\ell\to\infty$ (see App.~\ref{app:negbinom}).
This is consistent with our results on $P(\alpha)$ and shows likewise that the $\ell$-PSS alone does not converge to a coherent state.

\section{Autocorrelation functions and pseudo-lasing states.}
\label{sec:autocorr}
When we look at the autocorrelation functions $g^{(m)}_\ell$ [see Fig.~\ref{fig:numericgmpn}(a) and the boxes in Fig.~\ref{fig:poisson_subtract} for ${m=2,3,4}$] we make the somewhat contradictory observation that even though $P_{\ell}[n]$ does not converge to a Poisson distribution all $g^{(m)}_\ell$ approach the value one for increasing $\ell$.
In terms of the autocorrelation functions the $\ell$-PSSs of an initial thermal state reach the lasing regime since they meet one of the crucial criteria for lasing ($g^{(m)}=1$)
\cite{ulrich_photon_2007,chow_emission_2014} and we therefore call these states pseudo-lasing states.
To further investigate this behavior, we show that it is not limited to thermal states, but that the $g^{(m)}_\ell$ for a large class of initial states converge to one.
To this end we first calculate the $m$th factorial moment using the general expression for $P_\ell(\alpha)$ from Eq.~(\ref{eq:generalpss})
\begin{align}
\nn{m}_\ell = \int P_{0}(\alpha)\frac{\al^{2\ell}}{\nn{\ell}_0} \al^{2m}d\alpha^2=\frac{\mean{n^{(m+\ell)}}_0}{\mean{n^{(\ell)}}_0}.
\label{eq:facmoment}
\end{align}
We can now express the $m$th autocorrelation function of the $\ell$-PSSs in terms of the autocorrelation functions of the initial state
\begin{align}
  g^{(m)}_{\ell}=\frac{\nn{m}_\ell}{\n_\ell^m}=\frac{g^{(m+\ell)}_0}{g^{(\ell)}_0}
    \left (\frac{g^{(\ell)}_0}{g^{(1+\ell)}_0}\right )^m,
  \label{eq:gm_g0}
\end{align}
which was also derived in \cite{bogdanov_study_2016}.
For further analysis we express the next higher order autocorrelation function by a lower one, i.e.,~$g^{(\ell+1)}_0=c(\ell)g^{(\ell)}_0$,
with $c(\ell)$ being an arbitrary function of $\ell$.
In this formulation the autocorrelation of order $m+\ell$ reads
\begin{align}
  g^{(m+\ell)}_0&=\prod_{i=0}^{m-1}c(\ell+i)g^{(\ell)}_0
\end{align}
and $g^{(m)}_{\ell}$ can be expressed solely by quotients of $c(\ell)$
\begin{align}
  g^{(m)}_{\ell}&=\frac{\prod_{i=0}^{m-1}c(\ell+i) }{ c(\ell)^m}=\frac{c(\ell+m-1)\cdots c(\ell) }{ c(\ell)\cdots c(\ell)}.
  \label{eq:gm_cl}
\end{align}
The autocorrelations $g^{(m)}_{\ell}$ of the $\ell$-PSSs converge to one for every fixed  $m\in \mathbb{N}^+$ as $\ell\to\infty$ when
\begin{align}
  \lim_{\ell\to\infty}\frac{c(\ell+m)}{c(\ell)}=1
  \label{eq:ccondition}
\end{align}
holds.
This means in particular that all $c(\ell)$ (and thus all $g^{(m)}_0$) need to be non zero and cannot grow faster than a polynomial. For instance, thermal ($c(\ell)=\ell+1$) and coherent ($c(\ell)=1$) states meet this condition.
A counterexample is given by any initial state that can be represented by a finite number of Fock states.
In App.~\ref{app:condition}, we show that the condition in Eq.~(\ref{eq:ccondition}) is fulfilled for all $c(\ell)=\prod_{i=1}^{\ell}a(i)$ with ${\lim_{i\to\infty}a(i)=1}$.
We also show how, in principle, one can construct the photon number distribution \cite{barnett_statistics_2018} of an initial state that fulfills Eq.~(\ref{eq:ccondition}), i.e.,~a state that converges to a pseudo-lasing state.
We see that not only thermal but distributions from a large class of initial states converge to pseudo-lasing states with all $g^{(m)}\to 1$.
Note that in contrast to the lasing states the pseudo lasing states are only defined by the limiting behavior of their autocorrelation functions.
That means that they can still have quite different photon number distributions.

\subsection{Difference of the factorial moments}
\label{sec:diff}
The convergence of all autocorrelation functions to one seems to contradict the fact that the $\ell$-PSSs do not converge to a state with a Poissonian photon number distribution.
We can resolve this contradiction, by considering the difference between the factorial moments rather than their quotients ($g^{(m)}$).
We take the $m$th factorial moment of the $\ell$-PSSs [Eq.~(\ref{eq:facmoment})] and subtract the $m$th factorial moment of Poisson distribution
\begin{align}
\Delta^{(m)}_\ell= \nn{m}_\ell - \n_\ell^m,
\end{align}
i.e.,~the $m$th power of the corresponding intensity, which can be expressed solely by the $c(\ell)$ of the initial state
\begin{align}
  \frac{ \Delta^{(m)}_\ell}{\n_0^m} =
     \prod_{i=0}^{m-1}c(\ell+i)  -c(\ell)^m.
\end{align}
When we insert the $c(\ell)=\ell+1$ corresponding to a thermal state
\begin{align}
  \frac{ \Delta^{(m)}_\ell}{  \n_0^m}
    &= \left[\prod_{i=0}^{m-1}(\ell+1+i)\right]  -(\ell+1)^m\\
    &\ge  (m-1)\ell^{m-1},
  \label{eq:differences}
\end{align}
we see that difference of the factorial moments diverges in leading order with $\ell^{m-1}$, i.e.,~that the factorial moments of the $\ell$-PSS never match the ones of a lasing state. 

The order of divergence of $ \Delta^{(m)}_\ell$ resolves the paradox posed by the behavior of the pseudo-lasing states $(g^{(m)}_{\ell}\to1)$.
Since $\n_\ell^m$ is proportional to $\ell^{m}$, the denominator of $g^{(m)}_{\ell}$ grows just fast enough with $\ell$ to dominate the terms in its nominator proportional to $\ell^{m-1}$.
Therefore $\Delta^{(m)}_\ell$ diverges, while all $g^{(m)}_\ell$ converge to one.
This finding also showcases a possible downside of the (normalized) $g^{(m)}$ in its ability to monitor the lasing threshold.
These results show that there is no actual contradiction between the limiting behavior of $g^{(m)}_\ell$ and $P_\ell[n]$ and they are also consistent with our previous results on $\tilde{P}(\alpha)$ and $\tilde{P}[n]$. 
If we scale $\n^m_0$ according to Eq.~(\ref{eq:scaling}), we find that
$
 \tilde{\Delta}^{(m)}_\ell \propto \ell^{-1},
$
i.e.,~the difference vanishes for $\ell\to \infty$.

To further investigate the pseudo-lasing states we make a more general ansatz for the initial state characterized by $c(\ell)=(\ell+1)^{1-x}$ with $0\leq x\leq 1$ \footnote{All initial states characterized by $c(\ell)\sim\ell^b$ with $b>1$ converge to pseudo-lasing states, but all $\Delta^{(m)}_\ell$ diverge.}.
In this way we can monitor initial states "between" coherent ($x=1$) and thermal ($x=0$) states.
This ansatz means that we are multiplying Eq.~(\ref{eq:differences}) in leading order with the decaying sequence $\ell^{-mx}$
\begin{align}
  \label{eq:deltam}
  \Delta^{(m)}_\ell  \approx\n_0^m(m-1)\ell^{m(1-x)-1}.
\end{align}
We can conclude that for the subclass of pseudo-lasing states characterized by $c(\ell)=(\ell+1)^{\nicefrac{1}{m}}$ $(m'=\nicefrac{1}{(1-x)})$ not only all $g^{(m)}$ converge to one but also the first $m'-1$ factorial moments match the ones of the Poisson distribution exactly, while the differences of the higher order moments diverge.

\section{Conclusion}
\label{sec:concl}
So does the subtraction of multiple photons actually lead to a lasing state?
No, however, for a large class of initial states we can reach a pseudo-lasing state, i.e.,~a state with all autocorrelation functions converging to one while the photon number distribution is not Poissonian.
For a subclass of initial states, the first $m$ factorial moments of the pseudo-lasing states converge exactly to the ones of the Poisson distribution, while the others deviate unbounded.
The photon number distribution of these special pseudo-lasing states can provide a generalization to the compound Poisson distribution \cite{bogdanov_multiphoton_2017} interpolating between a thermal and a Poisson distribution.
A lasing state with an exact Poisson distribution can only be generated by photon subtraction, if the initial intensity is also scaled according to the number of subtractions.
Our results might foster the interpretation of recent experiments on (multi)photon subtracted states \cite{bogdanov_multiphoton_2017,
             avosopiants_non-gaussianity_2018,
             fedorov_quantum_2015,
             katamadze_how_2018,
             allevi_reliable_2010}
and demonstrate that $g^{(m)}\to 1$, for all $m$ is not equivalent to a convergence to a Poisson distribution \cite{gulyak_determination_2018}.

\acknowledgments
We are very grateful to I.~Carusotto for his comments on the manuscript and for fruitful discussions with U.~Peschel, S.~Barnett, and M.~Gerhold.
H.A.M.~Leymann acknowledges financial support from the European Union FET-Open grant MIR-BOSE 737017.

T.~Lettau and H.A.M.~Leymann have contributed equally to this work.


%

\appendix

\section{Auxiliary calculations}
\label{app:details}
\subsection{Continuous Poisson approximated by Gauss}
\label{app:continuos}
The continuous Poisson distribution
\begin{align}
  \Poisson_\ell(\lambda)=\frac{\lambda^\ell e^{(-\lambda)}}{n!}
\end{align}
has its maximum at $\lambda=\ell$.
In the following we show that this function can be approximated by a Gaussian for large $\ell$.
Since we are interested in the shape of this function for $\ell\gg 1$ in the neighborhood of its maximum, i.e.,~$\lambda=\ell(1+\delta)$ with $\delta\ll 1$,
we can use the sterling approximation $\ell!\approx \sqrt{2\pi \ell}e^{-\ell}\ell^\ell$.
If we insert the expressions we obtain
\begin{align}
  \Poisson_\ell(\lambda)
    &\approx\frac{[\ell(1+\delta)]^\ell e^{[-\ell(1+\delta)]}}{\sqrt{2\pi
      \ell}e^{-\ell}\ell^\ell}\\
    &=\frac{(1+\delta)^\ell e^{-\ell\delta}}{\sqrt{2\pi \ell}}.
  \label{eq:gauss_step2}
\end{align}
Now, we use $\delta\ll 1$ by taking the logarithm of $(1+\delta)^\ell$ and expanding it to $\ln(1+\delta)\approx \delta -\delta^2/2$.
Exponentiating this equation and inserting it into Eq.~(\ref{eq:gauss_step2}) leads to
\begin{align}
  \Poisson_\ell(\lambda)
    &\approx\frac{e^{\ell\delta-\ell\delta^2/2} e^{-\ell\delta}}{\sqrt{2\pi \ell}}=\frac{e^{-\ell\delta^2/2} }{\sqrt{2\pi \ell}}.
\end{align}
Replacing $\delta$ by $(\lambda -\ell)/\ell$ gives a Gaussian centered at $\ell$ with $\sigma^2=\ell$
\begin{align}
  \Poisson_\ell(\lambda)
    &\approx\frac{e^{-\frac{(\lambda-\ell)^2}{2\ell}} }{\sqrt{2\pi \ell}}.
\end{align}
For our purposes it is necessary to look at the Poisson distribution for $\lambda=(\ell+1)(1+\delta)$.
However, this does not change the argumentation above and we can substitute $\ell$ by $\ell+1$.

If we scale the intensity $\n_0=\n_{\textnormal{F}}/(\ell+1)$ [see Eq.~(\ref{eq:scaling})] an analog line of reasoning can be done to show that the scaled distribution $\tilde{P}_\ell(\alpha)$ can be approximated by Gaussian that is approaching a $\delta$-peak.
By differentiation we find that $\tilde{P}_\ell(\alpha)$ peaks at $\lambda=\ell/(\ell+1)$ so we consider it in the neighborhood of $\lambda=(1+\delta)\ell/(\ell+1)$
\begin{align}
  &\tilde{P}_{\ell}(\alpha)=\frac{(\ell+1)^{\ell+1}}{\pi\n_{\textnormal{F}} \ell!}
    e^{-\lambda(\ell+1)}\lambda^\ell\\
  &\approx\frac{(\ell+1)e^{-\ell\delta}e^{\ell\delta-\ell\delta^2/2}}{\pi\n_{\text{F}}
    \sqrt{2\pi \ell}} =
    \frac{e^{-\frac{(\al^2-\n_{\textnormal{F}})^2}{2\n_{\textnormal{F}}^2/(\ell+1)}}}
    {\pi \sqrt{2\pi \n_{\textnormal{F}}^2 /(\ell+1)}}.
\end{align}
\subsection{Ambiguity of the P-representation}
\label{app:uniquep}
Let us assume we have two different P-representations, $P_1(\alpha)$ and $P_2(\alpha)$, of the same state
\begin{align}
  \rho=\int P_{1,2}(\alpha)\ketbra{\alpha}{\alpha}d^2\alpha.
\end{align}
Using the optical equivalence theorem \cite{mandel_optical_1995} we can calculate any expectation value of an operator expressible in a normal ordered power series of creation and annihilation operators $g_N(a,a^\dagger)$ by integration $g(\alpha,\alpha^*)$ over the P-representations $P_{1,2}(\alpha)$
\begin{align}
  \textnormal{Tr}[\rho g_N(a,a^\dagger)]=\int P_{1,2}(\alpha)g(\alpha,\alpha^*)d^2\alpha.
  \label{eq:opticalequiv}
\end{align}
This means that for all $g(\alpha,\alpha^*)$ the integrals $\mean{g(\alpha,\alpha^*)}_{1,2}$ must be the same for both P-representations
\begin{align}
  0 &=\mean{g(\alpha,\alpha^*)}_{1}-\mean{g(\alpha,\alpha^*)}_{2}\nonumber\\
    &=\int (P_{1}(\alpha)-P_{2}(\alpha))g(\alpha,\alpha^*)d^2\alpha.
  \label{eq:equivzero}
\end{align}
When both representations are positive and only have a finite number of singularity's that are not stronger than a Dirac-peak,
we can always find regions $\Omega_>$ and $\Omega_<$ where $P_{1}(\alpha)>P_{2}(\alpha)$ and $P_{1}(\alpha)<P_{2}(\alpha)$, respectively.
Note that this is due to the normalization of $P(\alpha)$.
Thus the previous integral can be separated into two parts
\begin{align}
  I_{>/<}(g)=\int_{\Omega_{>/<}} (P_{1}(\alpha)-P_{2}(\alpha))g(\alpha,\alpha^*)d^2\alpha.
\end{align}
Let us now assume we have a power series $g'$ for which $I_{>}(g')=-I_{<}(g')$ and thus Eq.~(\ref{eq:equivzero}) holds.
In this case we can construct a second power series $f$ which fulfills the following conditions $f\geq/\leq 1$ in $\Omega_{>/<}$ with at least one measurable region in $\Omega_{>/<}$ where $f>/< 1$ holds.
The new function $fg'$ is still a valid test function and in this case $I_{>}(g')\neq-I_{<}(g')$.

We can conclude that for well behaved P-functions, as we discuss them in the main text, two different P-representations necessarily correspond to two different states,
since they do not result in the same expectation values.

\subsection{Calculation of the photon number distribution}
\label{app:calculation}

To obtain the photon number distribution from Eq.~(\ref{eq:poisson_p}) we need to solve the integral
\begin{align}
P_{\ell}[n]=\int P_{\ell}(\alpha) \frac{\al^{2n}}{n!}\exp(-\al^2) d^2\alpha,
\end{align}
which can be done by using the standard integral
\begin{align}
\pi^{-1}\int \exp(-C\al^2)\al^{2k}d^2\alpha=C^{-(k+1)}k!
\label{eq:standintegral}
\end{align}
and identifying $C=1+\n_0$ and $k=n+\ell$.
This results after some algebra in the expression for $P_{\ell}[n]$ given in Eq.~(\ref{eq:negbinom}).

\subsection{Negative binomial distribution converges to Poisson}
\label{app:negbinom}
When we scale the initial intensity in the photon number distribution [Eq.~(\ref{eq:negbinom})] according to Eq.~(\ref{eq:scaling}),
we obtain the compound Poisson distribution as reported in \cite{bogdanov_multiphoton_2017}
\begin{align}
\tilde{P}_{\ell}[n]=\frac{(\ell+n)! }{\ell!(\ell+1)^n}\,\frac{\n_{\textnormal{F}}^n}{ n!} \,\left(\frac{1}{1+\frac{\n_{\textnormal{F}}}{\ell+1}}\right)^{n+\ell+1},
\label{eqapp:compoundpoisson}
\end{align}
with a fixed mean $\mu=\n_{\textnormal{F}}$ and the coherence parameter $a=\ell+1$.
We can convince ourselves that this expression converges to a Poisson distribution for $\ell\to\infty$, since the first factor in Eq.~(\ref{eqapp:compoundpoisson}) converges to $1$ and the third converges to the exponential function $\exp(-\n_{\textnormal{F}})$.
Only with this scaled initial intensity the coherence parameter $a$ of the compound Poisson distribution can be increased independently of the mean $\mu$, which is necessary to facilitate the convergence to the regular Poisson distribution (compare Ref.~\cite{bogdanov_multiphoton_2017}).

\subsection{Condition for the autocorrelations of the initial state to reach the pseudo-lasing state}
\label{app:condition}

The general condition for the $c(\ell)$ of the initial state formulated in Eq.~(\ref{eq:ccondition}) is equivalent to the case $m=1$, since
\begin{align}
\lim_{\ell\to\infty}\frac{c(\ell+m)}{c(\ell)}=&\lim_{\ell\to\infty}\frac{c(\ell+m)}{c(\ell+m-1)}\cdots\frac{c(\ell+1)}{c(\ell)}\\\nonumber
=&\lim_{\ell\to\infty}\left(\frac{c(\ell+1)}{c(\ell)}\right)^m=1.
  \label{eq:newccondition}
\end{align}
All $c(\ell)$ that can be written in the form
\begin{align}
  c(\ell)=\prod_{i=1}^{\ell}a(i),\textnormal{ with }\lim_{i\to\infty}a(i)=1,
\end{align}
fulfill this condition [e.g., coherent: $a(i)=1$; thermal: $a(i)=(i+1)/i$].

\subsection{Reconstruction of the photon number distribution from the autocorrelation functions}
\label{app:reconstruct}
The generating function for the factorial moments of a discrete probability distribution 
\begin{align}
  M(\mu)= \sum_{m=0}^{\infty} (1-\mu)^m P[m]
  \label{eq:mom_gen_func}
\end{align}
can be expressed in the form \cite{barnett_statistics_2018}
\begin{align}
  M(\mu)=\sum_m \frac{(-\mu)^m}{m!}\nn{m}=\sum_m \frac{g^{(m)}(-\n\mu)^m}{m!},
  \label{eq:mom_gen_func2}
\end{align}
from which the probability distribution can be derived directly from the $g^{(m)}$, i.e. the $c(\ell)$
\begin{align}
  P[n]=n!^{-1}(-d_{\mu})^nM(\mu)|_{\mu=1}.
  \label{eq:dist:from_mom}
\end{align}
We need the $n$th derivative of a polynomial which is given by $(d_{\mu})^n \mu^m=\frac{m!}{(m-n)!}\mu^{m-n}$ for $m\geq n$.
Thus the photon number distribution can be written as
\begin{align}
  \label{eq:reconstruct}
  P[n] &=\sum_m n!^{-1}(-1)^n  \frac{g^{(m)}(-\n)^m}{m!}\frac{m!}{(m-n)!}\mu^{m-n} \Big|_{\mu=1}\nonumber\\
  &= \frac{\mean{n}^n}{n!}\sum_{m=0}^\infty \frac{g^{(m+n)}(-\mean{n})^{m}}{m!}.
\end{align}
Note that not all sets of $g^{(m)}$ are related to a positive normalized $P[n]$ which can be interpreted as a photon number distribution. 
Furthermore, while Eqs.~(\ref{eq:mom_gen_func}) and (\ref{eq:dist:from_mom}) hold in general, Eq.~(\ref{eq:mom_gen_func2}) will not converge for every photon number distribution \cite{barnett_statistics_2018}.

\section{Special pseudo-lasing states}

The analysis of the difference between the factorial moments of the pseudo-lasing states and a coherent state characterized by ${c(\ell)=(\ell+1)^{1/m}}$ [see  Eq.~(\ref{eq:deltam})] has shown that these ${c(\ell) }$ generate \emph{special} pseudo-lasing states,
for which all $\Delta^{(m'-1)}_\ell$ converge to zero for $\ell\rightarrow\infty$ and $m'<m$.
These special pseudo-lasing states have autocorrelation functions related to the thermal distribution $g^{(m)}_0 = (n!)^{\nicefrac{1}{m}}$ and lead to distributions with an interesting property:
For $m=\infty$ the distribution becomes the Poisson distribution whereas for $m=1$ it is the thermal distribution.
Similar to the compound distribution $m$ can be considered as a coherence parameter.
We can utilize Eq.~(\ref{eq:reconstruct}) to construct a power series  $({P[n]=\sum d_i \n^i})$ for the photon number distribution
\begin{align}
  P[n] = \frac{\mean{n}^n(n!)^{\nicefrac{1}{m}}}{n!}\sum_{i=0}^\infty
    \left (\frac{(i+n)!}{n!} \right)^{\nicefrac{1}{m}} \frac{(-\mean{n})^{i}}{i!}.
\end{align}
The radius of convergence of this series
\begin{align}
 \n_r &= \lim_{i\rightarrow \infty} \left| \frac{d_i}{d_{i+1}}\right|
     = \lim_{i\rightarrow \infty}\frac{i+1}{(i+n+1)^{1/m}} \\
    &= \lim_{i\rightarrow \infty}(i+n+1)^{1-1/m} = \infty,\,\text{ for } m\in [2, \infty) \nonumber
\end{align}
is unbounded for all $m\in [2, \infty)$.
Note that this does not ensure that the resulting photon number distribution of the special pseudo-lasing states is positive for all parameters.

\section{Delay-time dynamics of the second order autocorrelation function}
\label{app:delaytime}

The behavior of the photon autocorrelation function 
\begin{align*}
  g^{(2)}(t,\tau)
    = \frac{\mean{a^{\dagger}(t)a^{\dagger}(t+\tau)a(t+\tau)a(t)}}
           {\mean{a^{\dagger}(t)a(t)}^2}
\end{align*}
with respect to the delay time $\tau$ is determined by the dynamics of the system, i.e., by its Hamiltonian and its coupling to the environment.

When we, e.g., couple a $\ell$-PSS to a resonant cavity, the delay-time dynamics of $g^{(2)}(t,\tau)$ is governed by the Hamiltonian $\mathcal{H}=\hbar\omega a^{\dagger}a$ and the cavity losses to a thermal reservoir with mean photon number $\overline{n}$ and a loss rate $\kappa$. 
Employing the Lindblad formalism and the quantum regression theorem, we see that the $\tau$-dependence directly after the preparation of the $\ell$-PSS ($t=0$) is given by \cite{carmichael_statistical_1999}
\begin{align}
g^{(2)}(\tau)&=g^{(2)}_{\mathrm{PSS}}e^{-2\kappa \tau}+\frac{\overline{n}}{\n_{\mathrm{PSS}}}(1-e^{-2\kappa \tau}).
\end{align}
Here $g^{(2)}_{\mathrm{PSS}}$ and $\n_{\mathrm{PSS}}$ are the initial autocorrelation and intensity of the $\ell$-PSS at $t,\tau=0$.
The $\tau$-decay is solely determined by the cavity loss rate $\kappa$ and independent of the number of photon subtractions $\ell$.
On the other hand, if we analyze a laser model presented in \cite{carmichael_statistical_1999}, the autocorrelation of a laser field that is created in a cavity with a photon loss rate $\kappa$, has the following $\tau$-dependence 
\begin{align}
g^{(2)}(\tau)_{\mathrm{Laser}}&=1+\frac{1}{\n_{\mathrm{Sat}}(p-1)^2}(1-e^{-2\kappa(p-1) \tau}).
\end{align}
In this model the decay rate also depends on the pumping strength $p$.

We see that generically the $\tau$-dynamics ($g^{(2)}(\tau)$) of a 'conventionally' created laser field depends on the properties of the system, specifically on the parameter that controls the intensity and coherence, in this case the pump $p$. 
In contrast, the $\tau$-dynamics of a $\ell$-PSS is independent of the control parameter $\ell$. In this way one could demarcate a 'conventionally' created laser field from its photon subtracted counterpart by the decay rate of the respective autocorrelation functions.
The equations for the autocorrelation functions are derived in chapters 1.5.3 and 8.3.3 in \cite{carmichael_statistical_1999}.

\end{document}